# Resonant response of forced complex networks: the role of topological disorder

Hanshuang Chen[1], Yu Shen[1], Zhonghuai Hou[1,2,*], and Houwen Xin[1]

[1]*Department of Chemical Physics,* [2]*Hefei National Laboratory for Physical Sciences at Microscale,University of Science and Technology of China, Hefei, Anhui, 230026, People's Republic of China*

Abstract

We investigate the effect of topological disorder on a system of forced threshold elements, where each element is arranged on top of complex heterogeneous networks. Numerical results indicate that the response of the system to a weak signal can be amplified at an intermediate level of topological disorder, thus indicating the occurrence of topological-disorder-induced resonance. Using mean field method, we obtain an analytical understanding of the resonant phenomenon by deriving the effective potential of the system. Our findings might provide further insight into the role of network topology in signal amplification in biological networks.

**Many social, biological, and physical phenomena have been well understood on the topology of complex networks. Wide studies of complex networks have shown that dynamics taking place in heterogeneous networks is dramatically different from that in homogenous networks, such as epidemic spreading, synchronization process, percolation. In view of this, it is necessary that a new type of topological disorder should be introduced. However, there is lack of unique definition of topological disorder at present. Alternatively, we propose a definition of topological disorder via measuring the disparity of degree distribution. We numerically and analytically demonstrate an interesting phenomenon, topological-disorder-induced resonance, by using coupling forced threshold elements on complex networks. Our results might have potential importance in understanding the role of network topology in the process of signal amplification of real biological networks, such as neural network.**

*Electronic address: hzhlj@ustc.edu.cn

## Introduction

It is well-known that many social, biological, and physical systems can be properly described by complex networks whose nodes represent dynamical individuals and links mimic the interactions among them [1]. The study of complex networks has indicated that many real-world networks exhibit small-world [2] and scale-free [3] features that are neither regular nor completely random. On the other hand, many scholars have laid their attentions to dynamical process happening on top of complex networks. Specially, vast researches in this field have shown that dynamical behavior on heterogeneous networks is dramatically different from that on homogeneous networks [4]. For instance, topological disorder could lead to a vanishing percolation threshold [5], the whole infection of disease with any small spreading rate [6], the Ising model to be ordered at all temperatures [7], the transition from order to disorder in voter models [8], synchronization to be suppressed in oscillator network [9], spatiotemporal chaos to be tamed [10], etc [11].

In such complex systems, there exists an important kind of disorder, namely topological disorder. Although a great many investigations have recognized the importance of this kind of disorder, there is lack of unique definition so far. In fact, one of the most major sources of topological disorder is the disparity of node degree, where node degree is the number of edges connecting with the node. Therefore, it is feasible to characterize the topological disorder by standard deviation of node degree, as an alternative definition of the topological disorder. Disorder sometimes plays a counterintuitive role, for example, noise sometimes changes its role of conventional nuisance to a benefit. This is well-known as stochastic resonance (SR) [12], in which the right amount of noise is able to make a nonlinear dynamical system behave more regularly. Seminal works within the context of SR have been related to a weak forced bistable system [13] or excitable media [14] together with noise. When referring to the topological disorder, it is natural to ask whether the topological disorder induces a resonant behavior. To our best knowledge, it has not been reported previously in the literature. To the end, we here consider a system of coupled threshold elements, where the state of each element takes a binary value 0 or 1, which is simultaneously decided by the states of its neighbors as well as an external signal. The system is modeled via networked structure where nodes denote threshold elements and links connect them. We find that collective response of the system to the weak signal can be amplified at an intermediate range of the topological disorder, thus indicating the occurrence of topological-disorder-induced resonance.

The rest of this paper is organized as follows. In Section II, we propose the model and the approach to construct networks. Numerical simulation results of the model are presented in section III. Mean field analysis is put forward in section IV and main conclusions and discussion are addressed in section V.

**Model**

We consider a model of $N$ coupled elements on underlying networks, where the states of nodes $s_i = 0,1, i \in \{1,\cdots,N\}$ are binary. The state is updated according to the following rule [15]:

$$s_i(t+1) = \begin{cases} \Theta\left(\sum_{j\in\Omega_i} s_j(t) - Kh\right), & w.p. \quad 1-|f(t)| \\ \Theta\left(f(t)\right), & w.p. \quad |f(t)| \end{cases} \tag{1}$$

where w.p. denotes 'with probability', $\Omega_i$ is the set of neighbors of node $i$, and $\Theta(\cdot)$ is Heaviside step function. $f(t) = A\sin\omega t$ is the input signal, where $A$ and $\omega$ are the amplitude and frequency of the signal, respectively. $h$ is the threshold and $K$ is average degree. The model is a rather general paradigm for many real systems. The two states can be interpreted as being in favor an opinion or not, a neuron being firing or not, a gene being expressed or not, or several others.

We first construct a network where node degree follows Gaussian distribution. The network is generated according to the Molloy-Reed (MR) model [17]: each node is assigned a random number of stubs k that is drawn from a specified degree distribution. Pairs of unlinked stubs are then randomly joined. This construction eliminates degree correlations between neighboring nodes. We ensure that the average degree K keeps constant and let the standard deviation of degree $\sigma_g$ as the measure of topological disorder. Initially, each node is randomly assigned a state s(0)=0 or s(0)=1 with equal probability 1/2. We perform numerical calculations by Monte

Carlo (MC) simulation. At each run, the first $10^4$ time steps are discarded to achieve steady state and the following $10^4$ time steps are used to investigate the system's dynamics.

## Numerical Simulation

In Fig.1, we plot time evolution of the mean field $m(t)=N^{-1}\sum_{i=1}^{N}s_i(t)$ for different $\sigma_g$, with relevant parameters $N=1000$, $K=20$, $h=0.5$, $A=0.28$ and $\omega=0.05$. One can notice that m oscillates with different fashions for different $\sigma_g$. For $\sigma_g=0$ or relatively small values of $\sigma_g$, $m$ oscillates around the value one (or zero when the initial value $m(0)<0.465$) with the amplitude close to that of the input signal $A$. As $\sigma_g$ increases, e.g. for $\sigma_g=6$, $m$ oscillates nearly in the whole allowable range [0, 1] so as to dramatically increase the amplitude of the oscillations. With $\sigma_g$ increasing again, $m$ oscillates around a certain center value irrespective of $m(0)$, and both the amplitude and center value decreases as $\sigma_g$ increases.

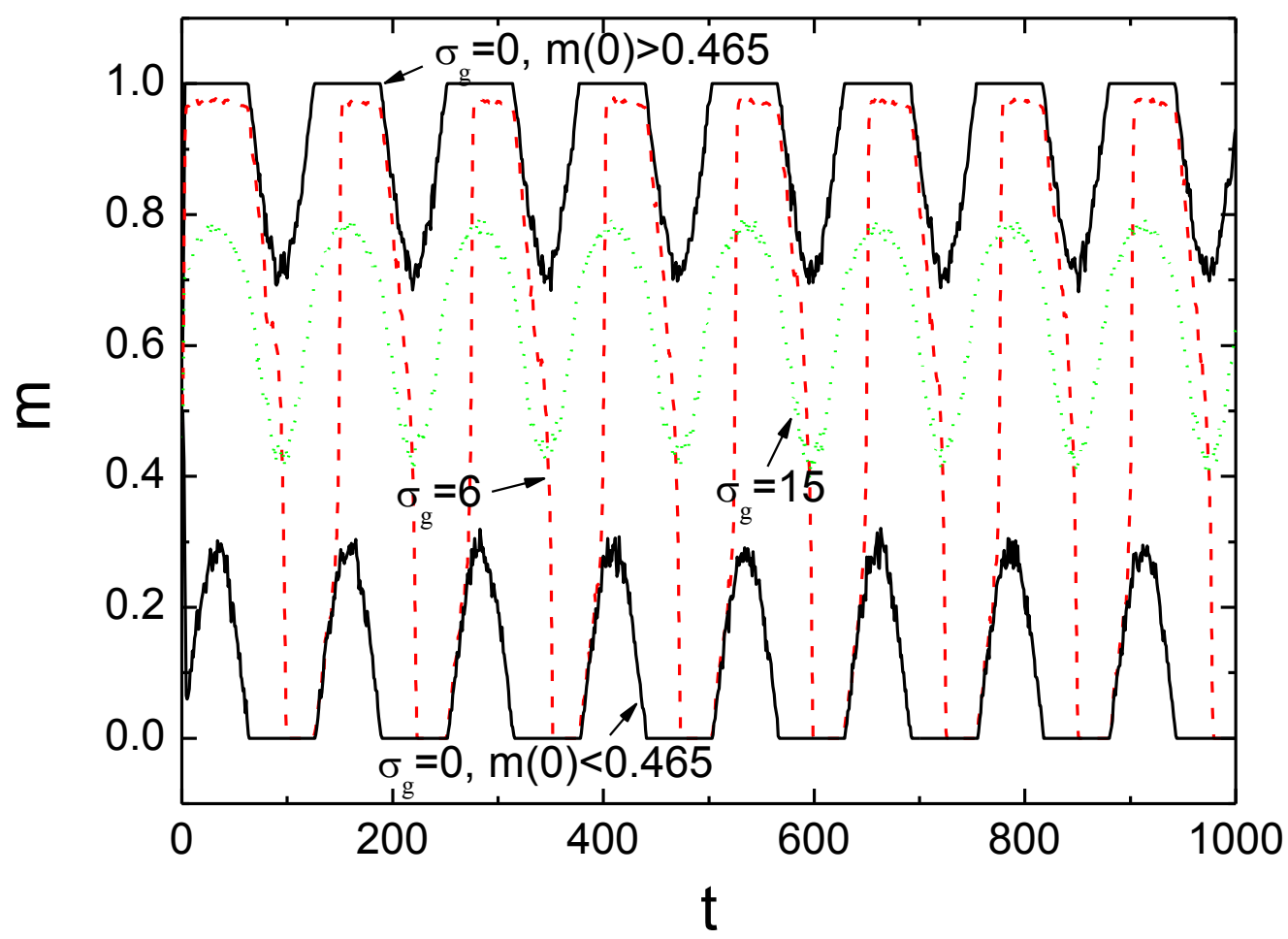

FIG. 1: Time evolution of mean field $m$ for different $\sigma_g$ on heterogeneous networks with Gaussian degree distribution. The other parameters are $N=1000$, $K=20$, $h=0.5$, $A=0.28$ and $\omega=0.05$

To quantify the response of the system to the input signal, we calculate the spectral amplification factor $R$, defined as the ratio of the output to input power at the corresponding driving frequency [16]

$$R = \left\langle \frac{4}{A^2} \left| \left\langle \left\langle m(t) e^{-i\omega t} \right\rangle \right\rangle \right| \right\rangle, \tag{2}$$

where $\langle\langle \square \rangle\rangle$ and $\langle \square \rangle$ denote average over time and initial conditions, respectively. The dependence of $R$ on $\sigma_g$ for different $A$ is shown in Fig.2, with relevant parameters $N = 1000$, $K = 20$, $h = 0.5$, $A = 0.28$ and $\omega = 0.05$. Here the amplitude of the input signal is set to be subthreshold, i.e. without topological disorder the response of the system to the signal is very faint. Each data is obtained via averaging over 50 different initial conditions and network realizations. With an increment of $\sigma_g$, $R$ reaches a maximum $R_C$ and then decreases, with $R_C$ corresponding to a moderate magnitude of topological disorder, and thus indicating the occurrence of topological-disorder-induced resonance. As the magnitude of the signal $A$ increases, the resonant peak becomes broader, the maximal height of the peak becomes lower, and the location of the peak slightly shifts to left. As the frequency of the signal $\omega$ increases, $R$ decreases monotonously and then vanishes when $\omega$ becomes rather large, as shown in the inset of Fig.2.

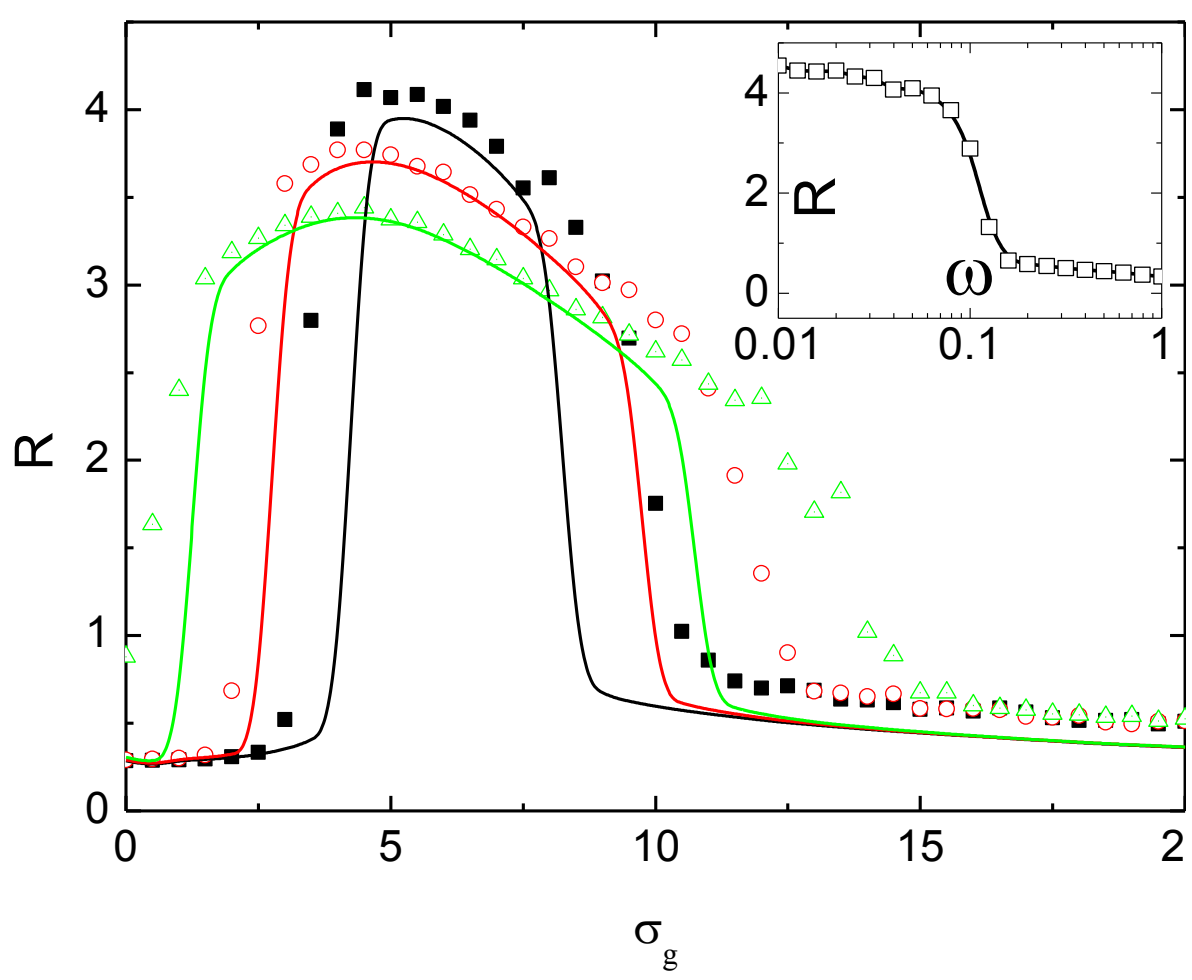

FIG. 2: $R$ as a function of $\sigma_g$ on heterogeneous network with Gaussian degree distribution for $A = 0.28$ (square), $A = 0.30$ (circle) and $A = 0.32$ (triangle), with relevant parameters $N = 1000$, $K = 20$, $h = 0.5$, and $\omega = 0.05$. Symbol and solid line corresponds to simulation result and analytical one, respectively. The inset depicts $R$ as a function of $\omega$ with $A = 0.28$ and $\sigma_g = 5$ fixed.

## Mean Field Analysis

In order to unveil the possible mechanism behind the above resonant phenomenon, we define $m_k$ as the probabilities that a node with degree $k$ is in state $s=1$, and $q$ as the probability that, for any node in the network, a randomly chosen nearest neighbor node is in state $s=1$. Furthermore, for any node the probability that a randomly chosen nearest neighbor node has degree $k$ is $kP_k/K$, where $P_k$ is degree distribution defined as the probability that a node chosen at random has degree $k$. It is supposed to be reasonable only in networks without degree correlation. The probabilities $m_k$ and $q$ satisfy the relation

$$q(t)=\sum_k kP_k m_k(t)/K. \qquad (3)$$

Note that $q(t)$ differs, in general, from $m(t)=\sum_k P_k m_k(t)$. In particular, for all nodes are in state $s=0$ or in state $s=1$, one has $q=m=0$ and $q=m=1$, respectively. Suppose $q$ is already known at a certain time $t$, one can calculate $m_k$ at the next time step. According to the evolution rule defined by Eq.1, in the absence of the external signal one has

$$m_k(t+1)=F(k,q)=\sum_{p=\lceil Kh \rceil}^{k} B(k,p,q), \qquad (4)$$

where $\lceil \square \rceil$ is ceiling function, and $B(k,p,q)=\frac{k!}{p!(k-p)!}q^p(1-q)^{k-p}$ is the binomial distribution. Thus, we insert Eq.4 to the right-hand side of Eq.3, which yields the evolution equation of $q$

$$q(t+1)=\Psi_1\left(q(t)\right)=\sum_k kP_k F(k,q(t))/K. \qquad (5)$$

The evolution equation of $m$ is readily written as

$$m(t+1)=\Psi_2\left(q(t)\right)=\sum_k P_k F(k,q(t)). \qquad (6)$$

We now add the external signal to Eq.5 and Eq.6, which become

$$q,m(t+1)=\left(1-|f(t)|\right)\Psi_{1,2}\left(q(t)\right)+|f(t)|\,\Theta\left(\sin\omega t\right). \qquad (7)$$

We iterate Eq.7 and then calculate the spectral amplification factor $R$ as a function of $\sigma_g$, as indicated by line in Fig.2. It is clear that theoretical analysis well predicts the trends that $R$ changes with $\sigma_g$ and $A$.

From time evolution of $m$ (shown in Fig.1), there appears to be some clues which the system behaves bistable. Actually, by iterating Eq.6, we find that for initial value of the mean field $m_0 > m_b$, $m$ converges to one stable fixed point $m_{w1} = 0$, or else another one $m_{w2} > 0$, which may imply the system is of bistable. Also, the evolution equation (5) has two stable fixed points $q_{w1}$ and $q_{w2}$, and an unstable one $q_b$. In Fig.3(a), we show these fixed points as a function of $\sigma_g$. For any $\sigma_g$, one has $m_{w1} = q_{w1} \equiv 0$, $m_b = q_b$ and $m_{w2} \le q_{w2}$. On the other hand, as $\sigma_g$ is increased, $m_b$, $q_b$, $m_{w2}$ monotonously decrease, while $q_{w2}$ decreases and then slightly increases.

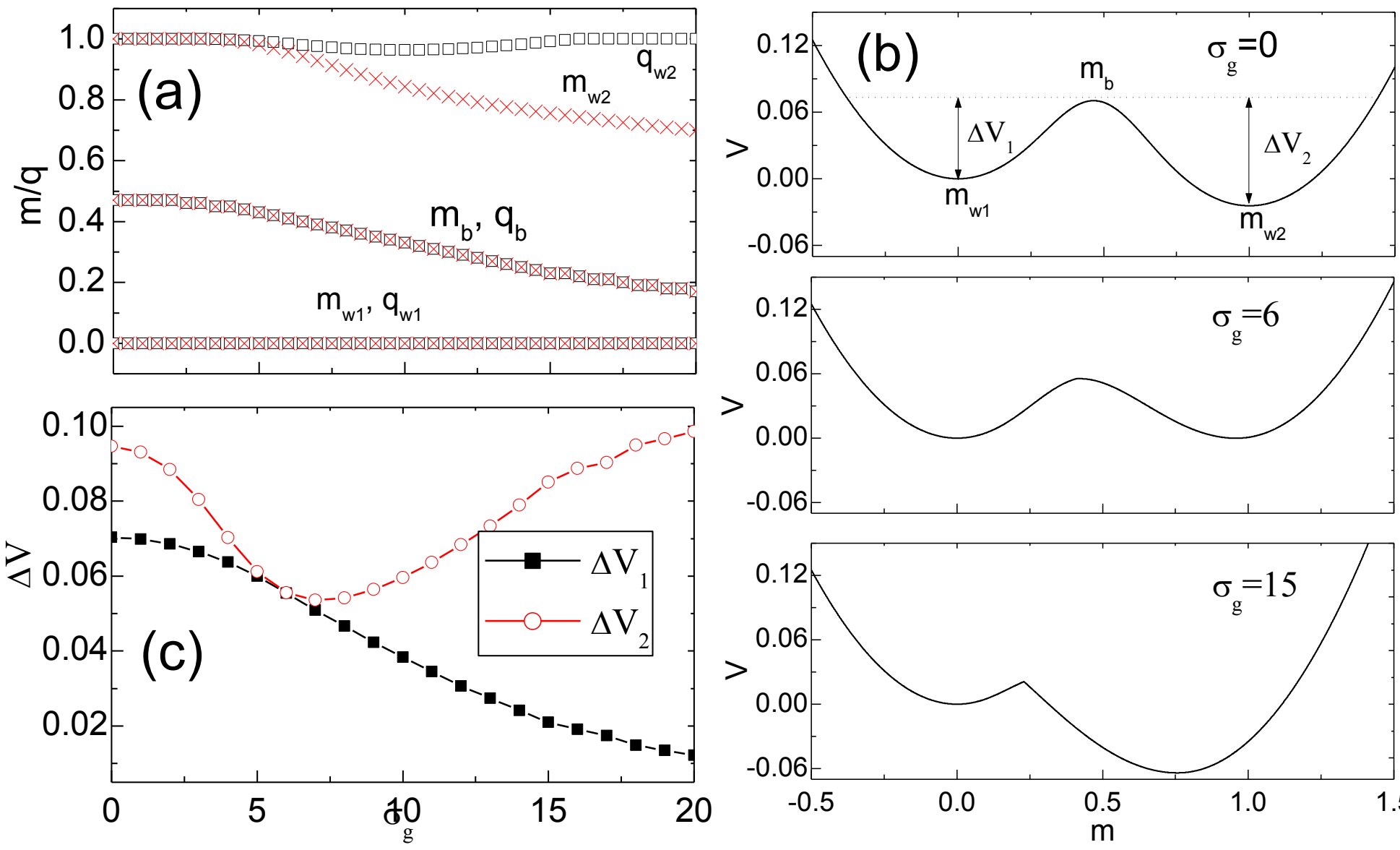

FIG. 3: (a) $m_b$ ($q_b$), $m_{w1}$ ($q_{w1}$) and $m_{w2}$ ($q_{w2}$) as a function of $\sigma_g$. (b) The effective potential of the system for three typical values of $\sigma_g$. (c) The height between potential wells and potential barrier $\Delta V$ (scheme in Fig.3(b)) as a function of $\sigma_g$.

It is interesting to visualize the effective potential $V(m)$ of the system, which can be done by converting the mean field equation to relaxational dynamics as $\frac{dm}{dt} = m(t+1) - m(t) = -\frac{\partial V(m)}{\partial m}$. For this purpose, we replace $q(t)$ in Eq.6 with $m(t)+\xi(t)$, where $\xi(t) = q(t) - m(t)$ can be calculated by Eq.5 and Eq.6. Thus, the effective potential $V(m)$ can be expressed as

$$V(m) = \frac{m^2}{2} - \int^m \sum_k P_k F\left(k, \lceil Kh \rceil, m' + \xi\right) dm'. \tag{8}$$

In Fig.3(b), the effective potentials for several typical values of $\sigma_g$ are shown, with relevant parameters $h = 0.5$ and $K = 20$. One can clearly distinguish that, for $\sigma_g = 0$ (homogeneous network) there exist two asymmetric potential wells $m_{w1}$, $m_{w2}$, separated by a potential barrier $m_b$. If we apply an external signal to the system and guarantee the signal is weak enough, $m$ will oscillates around one of the two potential wells (depending on the initial condition) with the amplitude comparable with that of the signal. As $\sigma_g$ is increased from $\sigma_g = 0$, the potential barrier and the potential well locating at $m_{w2}$ slightly shifts to left, as shown in Fig.3(a). Importantly, the potential barrier is lowered and the potential well locating at mw2 is raised, leading to the height between them, $\Delta V_1$ and $\Delta V_2$, decrease (shown in Fig.3(c)). Thus the system may surpass it from one potential well to another such that the response of the system to the weak signal is amplified. When $\sigma_g$ is further increased, $\Delta V_1$ still decreases but $\Delta V_2$ begins to increase. As a result, one potential well becomes very shallow but another one becomes very deep. With time evolution, the system quickly falls to the deeper potential well, and thus the system may fail to surpass the potential barrier again.

It is useful to consider an alternative type of degree distribution, e.g. uniform degree distribution in which node degree is randomly selected in the range $\left[K-\Delta, K+\Delta\right]$, where $\Delta$ is an integer between 0 and $K-1$. Similarly, the standard deviation $\sigma_u = \Delta/\sqrt{3}$ is the measurement of topological disorder. In Fig.4(a), we give simulation result and theoretical one of $R$ as a function of $\sigma_u$ for different $A$, which show excellent agreement between them. It is clearly shown that a maximum $R$ arises at an intermediate value of $\sigma_u$. No matter which type of network we use, topological-disorder always gives rise to a resonant response. Moreover, we also find in this case the system has bistable potential. As shown in Fig.4(b), we draw the bistable potential for three typical values of $\sigma_u$. Fig.4(c) depicts the height between the potential barrier and the two potential wells, $\Delta V_1$ and $\Delta V_2$, as a function of $\sigma_u$. The obtained results show that the underlying phenomena and mechanism in uniform-degree-distributed network are similar to those in Gaussian-degree-distributed network.

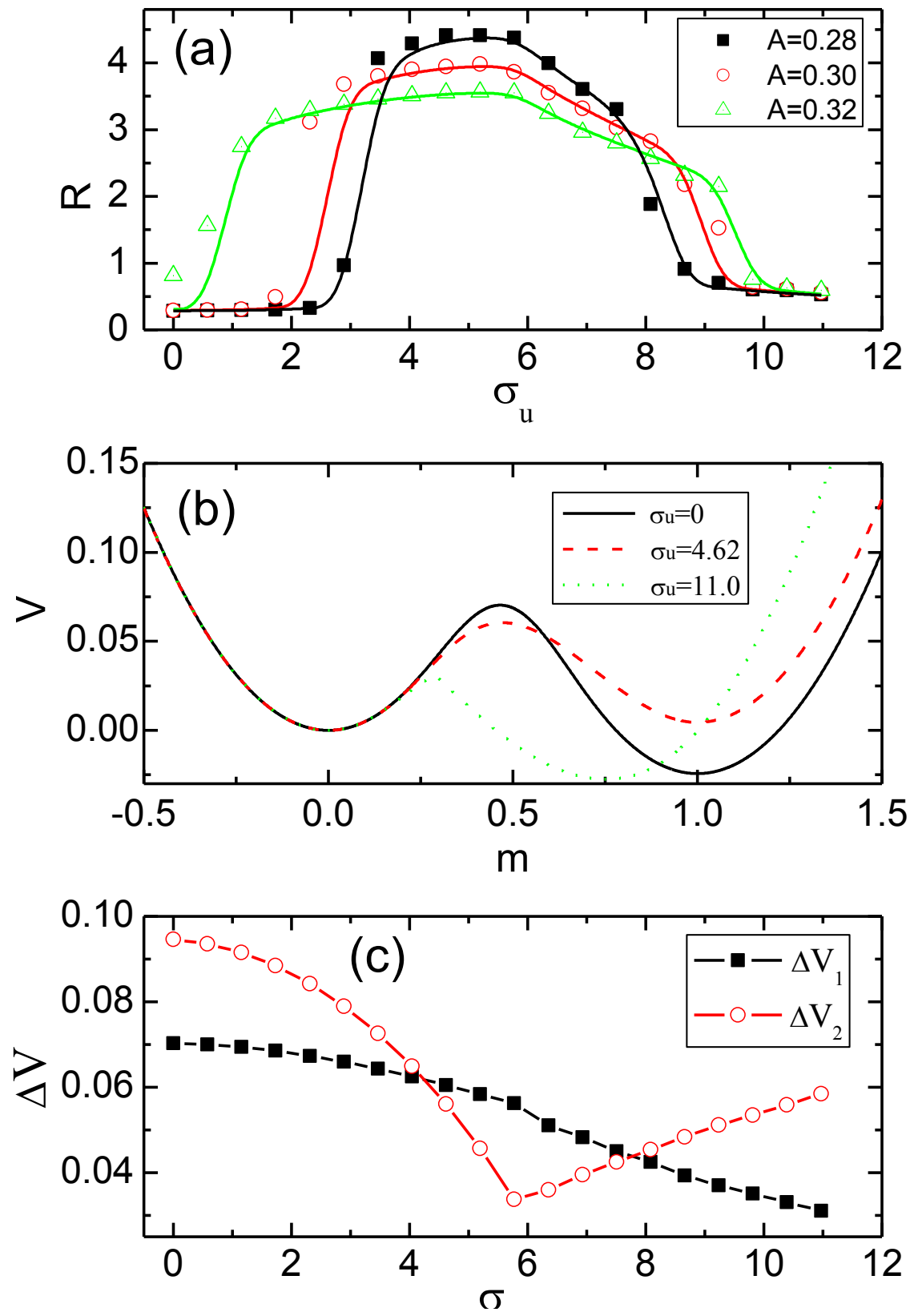

FIG. 4: (a) $R$ as a function of $\sigma_u$ on heterogeneous network with uniform degree distribution for different A. Symbol and line correspond to simulation result and theoretical prediction, respectively. (b) The effective potential of the system for three typical values of $\sigma_u$. (c) The height between potential wells and potential barrier $\Delta V$ (scheme in Fig.3(b)) as a function of $\sigma_u$. The other parameters are $N = 1000$, $K = 20$, $h = 0.5$, and $\omega = 0.05$.

## Consclusions and Discussion

In summary, we show the influence of topological disorder on the collective response of coupled threshold elements to a weak signal on complex network, where topological disorder exhibits the disparity of node degree on the underlying network. We find that the collective response to the weak signal can be amplified at a moderate level of topological disorder. We analytically formulate mean field equation, and find the effective potential of the system is bistable in which the potential has two asymmetric potential wells separated by a potential barrier. As topological disorder is increased, the height of one potential well is always decreases, while the height of another one is first decreases and then increases. Only at a moderate level of topological disorder, the system surpasses the potential barrier such that the system makes amplifying response to the signal. Thus we gain good understanding of topological-disorder-induced resonant behavior.

Topological disorder introduced in this paper stands for the disparity of node degree on the subject of complex networks, which can be easily distinguished itself from temporal disorder such as noise, and spatial disorder such as diversity in spatially extended systems [18, 19]. Although both noise [13, 14] and diversity [19] also induce a resonant behavior in bistable or excitable system, the starting point of our work is a different kind of disorder-topological disorder. Since the heterogeneity of degree is widespread in many biological network, our findings might give further understanding toward the influence of heterogeneous topological structure in the ability to amplify weak external signals.

Our numerical results and analytical treatment are based on the heterogeneous networks without degree-degree correlation. However, real-world networks often exhibit degree correlation feature [20]. Previous studies have shown that such degree mixing patterns have a significant influence in collective behavior of complex networks [20, 21]. Therefore, the effect of topological disorder in degree correlation networks on the present model deserves further investigation.

## Acknowledgments

This work was supported by the National Natural Science Foundation of China (projects no.

20673106).